\documentclass[12pt]{iopart}

\usepackage{amsopn}
\usepackage{amsfonts}
\usepackage{amssymb}
\usepackage{mathrsfs}
\usepackage{graphicx}
\usepackage{subfigure}

\DeclareMathOperator{\sech}{sech}
\DeclareMathOperator{\im}{Im}
\DeclareMathOperator{\re}{Re}

\begin{document}

\title[Goldstone modes of the NLS dark soliton]{Exact solutions to the four Goldstone modes  
around a dark soliton of the nonlinear Schr\"odinger equation}
\author{Andrew G. Sykes}
\address{Theoretical Division and Center for Nonlinear Studies, 
Los Alamos National Laboratory, Los Alamos, New Mexico 87545, USA}
\ead{sykes@lanl.gov}

\begin{abstract}
This article is concerned with the linearisation 
around a dark soliton solution of the nonlinear Schr\"odinger 
equation. Crucially, we present analytic expressions for the four linearly-independent zero eigenvalue solutions 
(also known as Goldstone modes) to the linearised problem. These solutions are then used to construct a 
Greens matrix which gives the first-order spatial response due to some perturbation. 
Finally we apply this Greens matrix to find the correction to the dark-soliton wavefunction of 
a Bose-Einstein condensate in the presence of fluctuations.
\end{abstract}

\submitto{\JPA}

\maketitle

\section{Introduction}

The nonlinear Schr\"odinger (NLS) equation is a ubiquitous nonlinear wave equation with a range of 
applications including the propagation of light within a 
waveguide~\cite{soliton_book_agrawal,soliton_book_hasegawa}, the behaviour 
of deep water waves~\cite{nls_deepwater}, and the mean-field theory of Bose-Einstein condensates~\cite{pitaevskii_becbook}. 
However, in many practical situations the NLS 
represents only the zeroth order approximation to the system, and for this reason, the response 
of an NLS system to small perturbations is important~\cite{kivshar_malomed}. A novel mathematical formalism (based on the 
four linearly independent Goldstone modes of the linearised problem)  
with which one can deal with the spatial consequences of such perturbations is the aim of this article. 
Currently relevant examples of such perturbative mechanisms include the loss 
and/or dephasing of coherent light traveling through an optical fibre, and the presence of quantum and/or 
thermal noise in Bose-Einstein condensates~\cite{franzeskakis}.

The problem under consideration has received a significant amount of attention in 
previous literature~\cite{keener_mclaughlin_soliton_perturbation1,
keener_mclaughlin_soliton_perturbation2,
kaup_newell_solitons,
herman_soliton_perturbation,
kaup_soliton_perturbation,
konotop_incorrect,
kivshar_soliton_perturbation}, and in fact it would seem that a general approach 
toward such problems has been established within the community since the late 1970's. 
Briefly, the approach focuses on finding eigenfunctions of a differential operator which is found from 
linearising the NLS around an analytic soliton solution. 
The majority of the earlier work was concerned more with the bright solitons 
found in the self-focusing NLS~\cite{zakharov_shabat1}. 
Progress on the dark soliton of the self-defocusing NLS caught up with its 
bright counterpart in the mid-to-late 1990's, with the introduction of a complete set of so-called 
``squared Jost solutions''~\cite{chinese_bdg_solutions}. 
The crux for the dark soliton solutions involved dealing with the nonvanishing boundary 
conditions, this issue was avoided in the earlier work of Ref.~\cite{konotop_incorrect}, where they 
force a vanishing boundary condition onto the perturbation for theoretical convenience. 
Reference~\cite{kivshar_soliton_perturbation} develops a method based on separating out the 
internal soliton dynamics from that of the boundary conditions, however such a separation 
is approximate at best~\cite{chinese_bdg_solutions,burtsev_camassa}. 
These squared Jost solutions of Ref.~\cite{chinese_bdg_solutions} elegantly provided the 
desired eigenfunctions for all real eigenvalues, except the case where the eigenvalue is zero 
(in this case the eigenfunctions are commonly referred to as the Goldstone modes). 
In this limit as the eigenvalue tends toward zero, the squared Jost solutions collapse down to just 
two linearly independent solutions (the linearised differential operator is ultimately a fourth order differential 
equation and should therefore yield four linearly independent solutions). 
This fact was noted in Refs.~\cite{chinese_bdg_solutions,bilaspavloff_darksoliton}, 
and two additional generalised eigenvectors 
were introduced to cope with the absence of the remaining two solutions. 
With the inclusion of these generalised eigenvectors it was shown 
that one had a complete set of functions. The main results of Ref.~\cite{chinese_bdg_solutions} has 
led onto several other publications~\cite{chinese_increment1,shengmeiao,chinese_greenfunction,chinesemultisolitons,
chinese_increment2} of a similar vein.

The issue has recently seen an influx of interest coming from the community of scientists involved with 
ultra-cold quantum gases. The original observation of dark-soliton-excitations of Bose-Einstein condensates 
within elongated trapping geometries came in 1999~\cite{darksoliton_first} and continues to accrue an impressive number 
of citations. Sophisticated numerical techniques have been employed in Refs.~\cite{ds_mishmash_carr2,ds_martin_ruostekoski,ds_martin_ruostekoski2} which 
investigate the lifetimes of dark solitons in the presence of quantum and thermal noise (special attention was paid 
to the high temperature regime in Ref.~\cite{ds_greeks_prl}). 
Analytic approaches toward the same problem were put forth earlier in Refs.~\cite{ds_dziarmaga1,ds_dziarmaga2}, 
and included the effects of the anomolous modes associated with phase diffusion of the Bose-Einstein condensate  
(originally considered in Ref.~\cite{bec_you_lewenstein}) as well as diffusion in the position of the dark soliton 
(these anomolous modes are given in equations~\ref{eq:om1}--\ref{eq:om2} of this article). Conflicting 
interpretations of the ensemble density evolution sparked debate as to whether the soliton exhibits decay or 
diffusion in the presence of noise~\cite{mishmash_comment}. Another mechanism put forth as being 
responsible for the decay of dark solitons is the effective three-body contact interactions 
considered in~\cite{ds_muryshev,ds_gangardt_kamenev}. They claim the soliton is protected against decay by the 
integrability in the system under two-body collisions. This integrability must be broken to observe soliton decay, 
a hypothesis which is supported by the claims of~\cite{ds_dziarmaga1,ds_dziarmaga2}.  The inclusion of 
three-body interactions destroys the integrability in the system. Further experiments 
in the field have successfully verified much of the fundamental interest in solitons such as their 
particle-like properties and mutual transparency under collision~\cite{ds_expt,ds_expt2}.

The overall goal of the present paper differs slightly from much of the previous literature, specifically we emphasise 
that time evolution of the soliton parameters is not addressed in this article (see Ref.~\cite{chinese_bdg_solutions,
chinese_greenfunction} 
for a treatment of this problem). Rather we concern ourselves solely with the first order correction to the spatial 
profile of the soliton. This correction is found by solving a nonhomogeneous fourth order differential 
equation (see equation \ref{eq:nonhomogeneous_equation} in section \ref{basic_formalism} of this paper). 
It is true that this correction can in principle be dealt with 
using the complete set of Ref.~\cite{chinese_bdg_solutions}, 
however this can be very difficult in general. 
Indeed it is expressly stated in Ref.~\cite{chinese_bdg_solutions} (see the final paragraph of the introduction) that 
the first order correction is difficult to obtain via their method. 
We present here a much simpler method based on analytic solutions for all four linearly 
independent Goldstone modes. It is the introduction of these analytic expressions for 
the two, previously unpublished, Goldstone modes which allows us to proceed in this way. 
The four solutions are related to the four fundamental symmetries of the NLS. These 
symmetries are phase symmetry, translational symmetry, Galilean symmetry, and 
dilaton symmetry. 
Aside from the methods aesthetic appeal, Ref.~\cite{me_dave_matt} describes a physical system 
(which necessitated the authors interest in this field), where, due to the numerical nature of the perturbing 
function, (denoted $g(x)$ in equation \ref{eq:nonhomogeneous_equation} of the current paper, but denoted $f(z)$ in 
Ref.~\cite{me_dave_matt}) the method of Ref.~\cite{chinese_bdg_solutions} was rendered useless.

The paper is organised as follows: In section \ref{basic_formalism} we set up the problem by 
linearising the NLS equation around a dark soliton. In section \ref{nonzero_eigenvalues} we look 
at the squared Jost solutions of Ref.~\cite{chinese_bdg_solutions} and discuss their importance as eigenfunctions 
of the linearised problem. In section \ref{zero_eigenvalues} we look at how these squared Jost solutions behave 
in the limit as the eigenvalue tends to zero. After establishing the fact that (in this zero eigenvalue limit) the 
squared Jost solutions give only two of the four possible eigenvectors, we give exact analytic solutions 
for all four eigenvectors. In section \ref{green_section} we use these eigenvectors to construct a Greens matrix for the 
differential operator of the linearised problem. In section \ref{example1} we illustrate the use of this Greens matrix in 
solving a practical example (specifically the correction to the dark-soliton wavefunction of a Bose-Einstein condensate, in the 
presence of fluctuations).

\section{Basic formalism}
\label{basic_formalism}

The usual nonlinear Schr\"odinger equation (with a defocusing nonlinearity), in its dimensionless form, is
\begin{equation}
-i\partial_t\psi-\frac{1}{2}\partial_z^2\psi+
|\psi|^2\psi=0,\label{eq:nls1}
\end{equation}
which, after a Galilean boost of the coordinates, ($x\equiv z-vt$) 
becomes
\begin{equation}
-i\partial_t\psi-\frac{1}{2}\partial_x^2\psi+iv\partial_x\psi+
|\psi|^2\psi=0.\label{eq:nls2}
\end{equation}
An interesting solution to equation \ref{eq:nls2} under 
non-vanishing boundary conditions 
is Tsuzuki's single soliton solution~\cite{tsuzuki1971,zakharov_shabat}. In this case, the function can be 
separated into the product 
$\psi(x,t)=e^{-it}\psi_0(x)$, (the Galiliean shift is important for this separation). 
The solution is then
\begin{equation}
\psi_0(x)=\cos(\theta)\tanh\left(x_c\right)+i\sin(\theta),
\label{eq:soliton}
\end{equation}
where $v=\sin(\theta)$ is the velocity of the soliton, and we have introduced a 
position coordinate $x_c=x\cos(\theta)$ for notational convenience. The boundary condition 
in use is 
$|\psi|\rightarrow1$ as $x\rightarrow\pm\infty$ (i.e. $|\psi_0|^2$ is normalised to unity 
far away from the soliton). 

Now let us consider a perturbation to this NLS system of the form; 
\begin{equation}
-i\partial_t\psi-\frac{1}{2}\partial_x^2\psi+iv\partial_x\psi+
|\psi|^2\psi=\epsilon F[\psi,\bar{\psi}].\label{eq:nls3}
\end{equation}
where $0<\epsilon\ll1$ and $F[\psi,\bar{\psi}]$ represents some process responsible 
for the departure from the ideal NLS and $\bar{\phantom{X}}$ denotes a complex conjugate. 
In a similar vein to Tsuzuki's solution of 
the unperturbed solution, we seek a separable solution in the form
\begin{equation}
\psi(x,t)=e^{-it}\left[\psi_0(x)+\epsilon\psi_1(x,T_0,T_1,\ldots)+\epsilon^2\psi_2(x,T_0,T_1,\ldots)\right]
\label{eq:expansion}
\end{equation}
where the coordinates $T_n=\epsilon^nt$, for $n=0,1,2,\ldots$, introduce a multiple-time-scale analysis. 
In the limit as $\epsilon\rightarrow0$ the coordinates $T_0,T_1,\ldots$ may be regarded as being independent.

As an aside, we note that a solution to equation \ref{eq:nls3} in the form of equation \ref{eq:expansion} is 
certainly not guaranteed, however the ansatz may be appropriate in certain scenarios. 
To aid any reader, who is interested in the application of this work, 
in determining whether or not the ansatz of equation \ref{eq:expansion} is appropriate in a particular 
case we outline a few basic points. 
\begin{itemize}
 \item When $\epsilon=0$ the system is a perfect NLS system and the function $\psi$ is given by 
Tsuzuki's single soliton solution. Changes in $\psi$ occur over a length scale $x_c\approx1$ and a time 
scale $t\approx1$.
 \item For finite $\epsilon$ the system will acquire an additional dynamical evolution which occurs over 
a timescale $\epsilon t\approx1$~\cite{kaup_newell_solitons}, aswell as a new spatial profile (given 
by the spatial dependence of $\psi_1$) which is an $O(\epsilon)$ correction to $\psi_0(x)$.
\end{itemize}

Continuing on with the formalism, we expand the time derivative as 
$\partial_t=\partial_{T_0}+\epsilon\partial_{T_1}+\ldots$ and look for a solution of $\psi_1$ under the 
assumption that the rapid-time evolution (if any exists) is complete, that is $\partial_{T_0}\psi_1=0$. 
Inserting equation \ref{eq:expansion} into equation \ref{eq:nls3} and keeping only the 
terms which are linear in $\epsilon$ we get 
\begin{equation}\fl
 \left[-\frac{1}{2}D_x^2+ivD_x+2|\psi_0|^2-1\right]\psi_1+\psi_0^2\bar{\psi}_1=
F\left[\psi_0e^{-it},\bar{\psi}_0e^{it}\right]e^{it},\label{eq:X1}
\end{equation}
where $D_\alpha\equiv\frac{d\phantom{\alpha}}{d\alpha}$. Crucially for this particular approach to be relevant, the right-hand-side 
of equation \ref{eq:X1} should not depend on the rapid-time variable $T_0$. 
The severity of this condition is unclear in general, however, at least in the case of one-dimensional 
Bose-Einstein condensates (where the author first encountered this kind of problem) 
this condition is certainly true. The problem then is finding a 
solution for the perturbation $\psi_1$. This is given by the following fourth order, nonhomogeneous 
differential equation; 
\begin{equation}
\mathcal{H}_x\left[\begin{array}{c}
\psi_1(x,T_1)\\ \bar{\psi}_1(x,T_1)
\end{array}\right]=
\left[\begin{array}{c}
g(x,T_1)\\ \bar{g}(x,T_1)
\end{array}\right]\label{eq:nonhomogeneous_equation}
\end{equation}
where 
\begin{equation}\fl
\mathcal{H}_x=\left[\begin{array}{cc}
-\frac{1}{2}D_x^2+ivD_x+2|\psi_0(x)|^2-1 & 
\psi_0(x)^2 \\
\bar{\psi}_0(x)^2 & 
-\frac{1}{2}D_x^2-ivD_x+2|\psi_0(x)|^2-1
\end{array}
\right].\label{eq:Hoperator}
\end{equation}
The function $g$ is the right hand side of equation \ref{eq:X1}, and can only depend 
on the slow-time variable $T_1$. We will refer to the linear operator $\mathcal{H}_x$ as the 
linearised operator. The eigenfunctions of this operator play an important part in the solution to 
equation \ref{eq:nonhomogeneous_equation}.

\section{Eigenfunctions of the linearised operator}

\subsection{Non-zero eigenvalues}
\label{nonzero_eigenvalues}
In this section we briefly review some previous literature on this problem~\cite{chinese_bdg_solutions,
chinese_greenfunction,chinese_increment1,chinese_increment2}. 
Specifically we look for solutions to 
\begin{equation}
 \mathcal{H}_x\left[\begin{array}{c}
u_E(x)\\ v_E(x)
\end{array}\right]=E\left[\begin{array}{c}
u_E(x)\\ -v_E(x)
\end{array}\right]\label{eq:eigenequation}
\end{equation}
for a fixed $E\neq0$. Four linearly independent functions $u^j_E$ and $v^j_E$ can be found 
by searching the previous literature~\cite{chinese_bdg_solutions},
\begin{eqnarray}
 u^j_{E}=e^{ik_jx}\left[k_j/2+E/k_j+i\cos(\theta)\tanh\left(x_c\right)\right]^2 \label{eq:bdg_solns1a}\\
 v^j_{E}=e^{ik_jx}\left[k_j/2-E/k_j+i\cos(\theta)\tanh\left(x_c\right)\right]^2 \label{eq:bdg_solns1b}
\end{eqnarray}
where $j=1,2,3,4$ and $k_j$ is one of the four roots to the polynomial $\left[E+k\sin(\theta)\right]^2=k^2(k^2/4+1)$. 
It is worth while to note that two of the roots ($k_1$ and $k_2$ say) are real, while 
two of the roots ($k_3$ and $k_4$ say) are complex. The complex roots mean $u^{3,4}_E$ and $v^{3,4}_E$ 
diverge exponentially as $x$ tends to either positive or negative infinity and for this reason are usually 
excluded on the grounds that they are unphysical.

Equations \ref{eq:bdg_solns1a}--\ref{eq:bdg_solns1b} can be thought of as the radiative eigenvectors of 
$\mathcal{H}_x$. Plane wave excitations moving through the system essentially see the dark soliton as a reflectionless 
potential and emerge on the other side with nothing more than a phase shift.

\subsection{Zero eigenvalues}
\label{zero_eigenvalues}

As well as the radiative eigenvectors of the previous subsection, one also has a discrete set of eigenvectors associated 
with the symmetries of equation~\ref{eq:nls1}. These are nonradiative eigenvectors and are commonly referred 
to as Goldstone modes. They have zero energy, but they have physical effects such as changing the phase of the 
soliton, shifting its spatial position, or dilating its profile. 
We thus turn our attention to solving the homogeneous problem,
\begin{equation}
 \mathcal{H}_x\left[\begin{array}{c}
\omega(x)\\ \bar{\omega}(x)
\end{array}\right]=\left[\begin{array}{c}
0\\ 0
\end{array}\right],\label{eq:zero}
\end{equation}
to find these Goldstone modes. 
The fact that equation \ref{eq:eigenequation} is solved for $E\neq0$ would seem to indicate that 
solutions to equation \ref{eq:zero} could be found simply by taking the limit $E\rightarrow0$. Unfortunately 
this isn't the case, as $E\rightarrow0$ the four solutions of equations \ref{eq:bdg_solns1a}--\ref{eq:bdg_solns1b} collapse down 
into just two linearly independent solutions,
\begin{eqnarray}
 \left[\begin{array}{c}
\omega_1(x)\\ \bar{\omega}_1(x)
\end{array}\right]&=
\left[\begin{array}{c}
i\left(\cos(\theta)\tanh\left(x_c\right)+i\sin(\theta)\right)\\ 
-i\left(\cos(\theta)\tanh\left(x_c\right)-i\sin(\theta)\right)
\end{array}\right]=
\left[\begin{array}{c}
i\psi_0\\ -i\bar{\psi}_0
\end{array}\right]\nonumber
\\
\left[\begin{array}{c}
\omega_2(x)\\ \bar{\omega}_2(x)
\end{array}\right]&=
\left[\begin{array}{c}
\sech^2\left(x_c\right)\\ 
\sech^2\left(x_c\right)
\end{array}\right]\nonumber
\end{eqnarray}
and so we find that two of the solutions are absent from the previous literature. This 
point has not gone unnoticed, and the usual strategy for dealing with these absent 
solutions is to find generalised eigenvectors which satisfy 
\begin{equation}
 \mathcal{H}_x\left[\begin{array}{c}
\Omega(x)\\ \bar{\Omega}(x)
\end{array}\right]=\left[\begin{array}{c}
\omega(x)\\ \bar{\omega}(x)
\end{array}\right].\label{eq:generalised}
\end{equation}
The previous literature contains expressions for two such generalised eigenvectors 
(see for example, appendix A of Ref.~\cite{bilaspavloff_darksoliton}) 
and it is the union of the $\mathcal{H}_x$ and $\mathcal{H}_x^2$ null-spaces which is then 
used to form a complete set of functions. 

Rather than adopt this approach based on generalised eigenvectors, we write down expressions 
for all four linearly independent solutions to equation \ref{eq:zero}
\begin{eqnarray}\fl
\omega_1(x)=-\sin(\theta)+i\cos(\theta)\tanh(x_c)
\label{eq:om1}\\ 
\fl
\omega_2(x)=\sech^2(x_c)\label{eq:om2}\\
\fl
\omega_3(x)=\sech^2(x_c)\left[2x_c-x_c\cosh(2x_c)+
(3/2)\sinh(2x_c)\right]\tan(\theta)
+\nonumber\\
2i\left[x_c\tanh(x_c)-1\right]\label{eq:om3}\\
\fl
\omega_4(x)=\sech^2(x_c)\left\{x_c\left(10-4\cos^2(\theta)-8\sin(\theta)
\sin(\theta-2ix_c)\right)+\right.\nonumber\\
\left.\cosh(x_c)\left[i\sin(2\theta-3ix_c)-5i\sin(2\theta-ix_c)\right]
+6\sinh(2x_c)\right\}\label{eq:om4}.
\end{eqnarray}
These four expressions form the key result of this paper ($\omega_1$ and $\omega_2$ have appeared in the 
previous literature, however to the best of our knowledge $\omega_3$ and $\omega_4$ have not). 
These expressions do not follow from the 
finite $E$ eigenvectors, rather they are related to the four fundamental symmetries of the NLS;
$\omega_1$ $\leftrightarrow$ phase symmetry, $\omega_2$ $\leftrightarrow$ translational symmetry, 
$\omega_3$ $\leftrightarrow$ Galilean symmetry, and $\omega_4$ $\leftrightarrow$ dilaton symmetry. 
A brief summary of these symmetries is given below: 
Assuming that $\phi_0(x,t)$ is a solution of equation \ref{eq:nls1} and $\alpha$ is any real constant, then
\begin{itemize}
 \item \emph{phase symmetry} tells us that $\phi_0'(x,t)\equiv e^{i\alpha}\phi_0(x,t)$ will also be a solution,
\item \emph{translational symmetry} tells us that $\phi_0'(x,t)\equiv \phi_0(x-\alpha,t)$ will also be a solution,
\item \emph{Galilean symmetry} tells us that $\phi_0'(x,t)\equiv e^{i(\alpha x-\frac{\alpha^2}{2}t)}\phi_0(x-\alpha t,t)$ will also be a solution,
\item \emph{dilaton symmetry} tells us that $\phi_0'(x,t)\equiv\alpha\phi_0(\alpha x,\alpha^2t)$ will also be a solution.
\end{itemize}

In order to show the linear independence of equations \ref{eq:om1}--\ref{eq:om4} we calculate the 
Wronskian 
\begin{equation}\fl
 \left|\begin{array}{cccc}
        \omega_1(x) & \omega_2(x) & \omega_3(x) & \omega_4(x) \\
        \omega_1'(x) & \omega_2'(x) & \omega_3'(x) & \omega_4'(x) \\
        \omega_1''(x) & \omega_2''(x) & \omega_3''(x) & \omega_4''(x) \\
        \omega_1'''(x) & \omega_2'''(x) & \omega_3'''(x) & \omega_4'''(x) \\ 
       \end{array}
\right|=512\cos^5(\theta)\sech^4(x_c)\sin^4(\theta-ix_c),\nonumber
\end{equation}
and we see that, provided $0\leq\theta<\pi/2$, the solutions are linearly independent. In the case where $\theta=\pi/2$ the soliton 
has vanished from the system and the problem becomes trivial.

\section{Constructing a Greens matrix}
\label{green_section}

Returning our attention to the solution of equation \ref{eq:nonhomogeneous_equation}, we 
use the zero-eigenvalue solutions given in equations \ref{eq:om1}--\ref{eq:om4} to construct a Greens matrix for the 
linearised operator. The minimum requirement for this Greens matrix being that it satisfies the following condition;
\begin{equation}
 \mathcal{H}_x\tilde{G}(x,s)=\mathbb{I}_2 \delta(x-s)\label{eq:green}
\end{equation}
where $\mathbb{I}_2$ is the $2\times2$ identity matrix and $\tilde{G}$ denotes the $2\times2$ Greens matrix. 
The general solution to equation \ref{eq:nonhomogeneous_equation} will then 
be given by 
\begin{equation}
 \left[\begin{array}{c}
\psi_1(x)\\ \bar{\psi}_1(x)
\end{array}\right]=\int_{-\infty}^\infty \tilde{G}(x,s)\left[\begin{array}{c}
g(s)\\ \bar{g}(s)
\end{array}\right]\label{general_solution}
\end{equation}
Additional requirements given by symmetry and boundary conditions of the specific problem will completely determine $\tilde{G}$. 

We write $\tilde{G}$ as,
\begin{equation}
 \tilde{G}(x,s)=\sum_{j=1}^4\left\{\begin{array}{ll}

\left[\begin{array}{c} \omega_j(x)\\ \bar{\omega}_j(x)\end{array}\right]
\left[\begin{array}{cc} \bar{\kappa}_j(s) & \kappa_j(s)\end{array}\right] & \quad s<x
\\
\left[\begin{array}{c} \omega_j(x)\\ \bar{\omega}_j(x)\end{array}\right]
\left[\begin{array}{cc} \bar{\lambda}_j(s) & \lambda_j(s)\end{array}\right] & \quad x<s
   
               \end{array}
\right.
\end{equation}
and equation \ref{eq:green} gives rise to the conditions when $x=s$,
\begin{eqnarray}
&\lim_{x\to s^+}\tilde{G}(x,s)=\lim_{x\to s^-}\tilde{G}(x,s)\\
&\left[\lim_{x\to s^+}D_x\tilde{G}(x,s)\right]-\left[\lim_{x\to s^-}D_x\tilde{G}(x,s)\right]=-2\mathbb{I}_2.
\end{eqnarray}
These conditions manifest in the following simultaneous equations for $\kappa_j$ and $\lambda_j$;
\begin{eqnarray}
\kappa_1(s)-\lambda_1(s)=&\frac{1}{2}\sec^2(\theta)\omega_3(s),\label{eq:bc1}\\
\kappa_2(s)-\lambda_2(s)=&\frac{1}{4}\sec(\theta)\tan(\theta)\omega_3(s)+\frac{1}{16}\sec^3(\theta)\omega_4(s),\label{eq:bc2}\\
\kappa_3(s)-\lambda_3(s)=&-\frac{1}{2}\sec^2(\theta)\omega_1(s)-\frac{1}{4}\sec(\theta)\tan(\theta)\omega_2(s),\label{eq:bc3}\\
\kappa_4(s)-\lambda_4(s)=&-\frac{1}{16}\sec^3(\theta)\omega_2(s).\label{eq:bc4}
\end{eqnarray}
The symmetry of $\tilde{G}$ [namely $\tilde{G}(x,s)=\tilde{G}^\dagger(s,x)$, where $\dagger$ denotes the complex conjugate] yields a further condition;
\begin{equation}
 \sum_{j=1}^4\bar{\lambda}_j(s)\omega_j(x)=\sum_{j=1}^4\kappa_j(x)\bar{\omega}_j(s).\label{eq:bc_sym}
\end{equation}
Because $\tilde{G}(x,s)$ must also be a solution to the adjoint problem $\tilde{G}(x,s)\mathcal{H}^\dagger_s=\mathbb{I}_2\delta(x-y)$ 
(where $\mathcal{H}_s^\dagger$ acts to the left), we see that $\kappa_j$ and $\lambda_j$ must be linear combinations 
of the $\omega_j$. Thus we look for 32 real constants, $\kappa_i^j$ and $\lambda_i^j$ (where $i,j=1,2,3,4$) which 
appropriately define 
\begin{eqnarray}
 \kappa_i(s)=\sum_{j=1}^4\kappa_i^j\omega_j(s),\\
 \lambda_i(s)=\sum_{j=1}^4\lambda_i^j\omega_j(s).
\end{eqnarray}
Equations \ref{eq:bc1}--\ref{eq:bc4} then become
\begin{eqnarray}
 \kappa_1^j-\lambda_1^j&=\delta_{j3}\frac{1}{2}\sec^2(\theta),\\
\kappa_2^j-\lambda_2^j&=\delta_{j3}\frac{1}{2}\sec^2(\theta)+\delta_{j4}\frac{1}{16}\sec^3(\theta),\\ 
\kappa_3^j-\lambda_3^j&=-\delta_{j1}\frac{1}{2}\sec^2(\theta)-\delta_{j2}\frac{1}{16}\sec^3(\theta),\\
\kappa_4^j-\lambda_4^j&=-\delta_{j2}\frac{1}{16}\sec^3(\theta),
\end{eqnarray}
(where $\delta_{jk}$ is the Kronecker delta) while equation \ref{eq:bc_sym} becomes
\begin{equation}\fl
 \lambda_2^1=\kappa_1^2,\quad
 \lambda_3^1=\kappa_1^3,\quad
 \lambda_3^2=\kappa_2^3,\quad
 \lambda_4^1=\kappa_1^4,\quad
 \lambda_4^2=\kappa_2^4,\quad
 \lambda_4^3=\kappa_3^4.
\end{equation}
We can also set $\lambda_1^1=\lambda_2^2=\lambda_3^3=\lambda_4^4=0$ since these diagonal elements only 
affect the final solution for $\psi_1(x)$ by adding a constant times $\omega_j(x)$ which is of no 
physical interest since it is just transforming the solution into one of the four previously-mentioned symmetry 
groups. This leaves us with 26 equations for the 32 unknowns, the remaining 6 equations are provided by the 
boundary conditions on $\psi_1(x)$.

\section{Example problem}

\subsection{1D Bose-Einstein condensate in the presence of fluctuations}
\label{example1}

\begin{figure}
\centering
\subfiguretopcaptrue
\subfigure[]{\includegraphics[width=6cm]{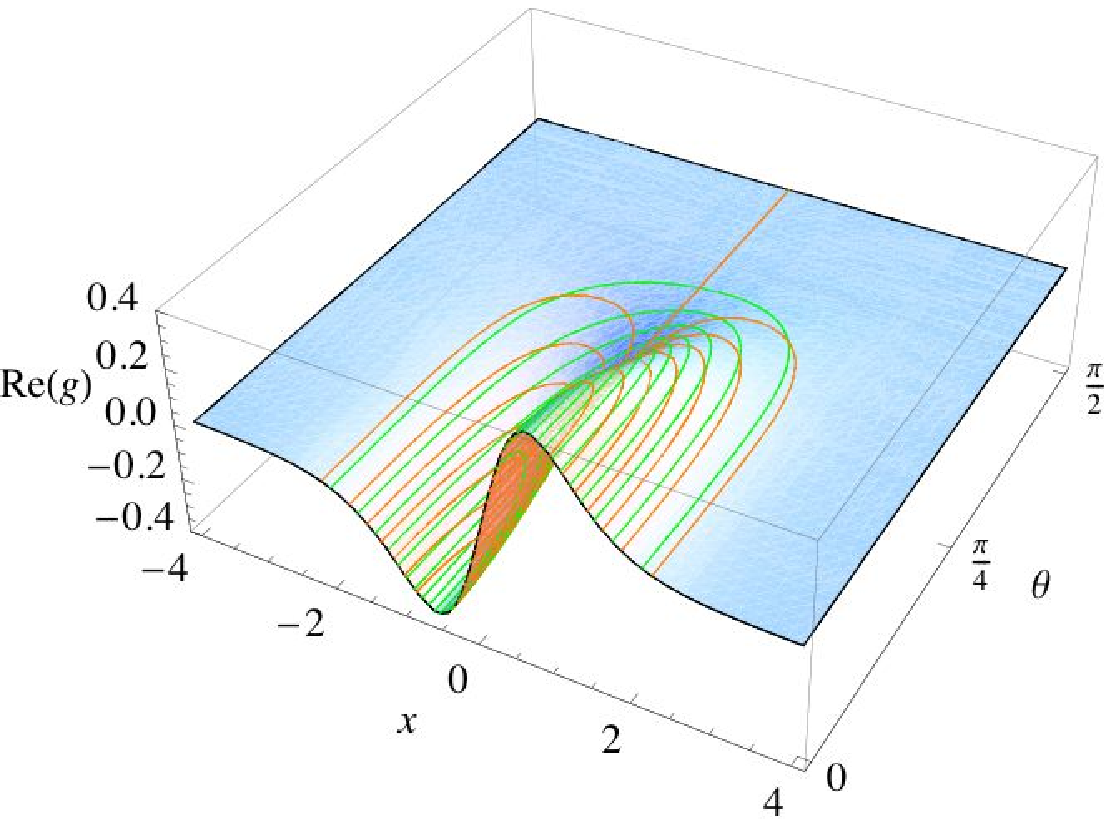}}
\qquad
\subfigure[]{\includegraphics[width=6cm]{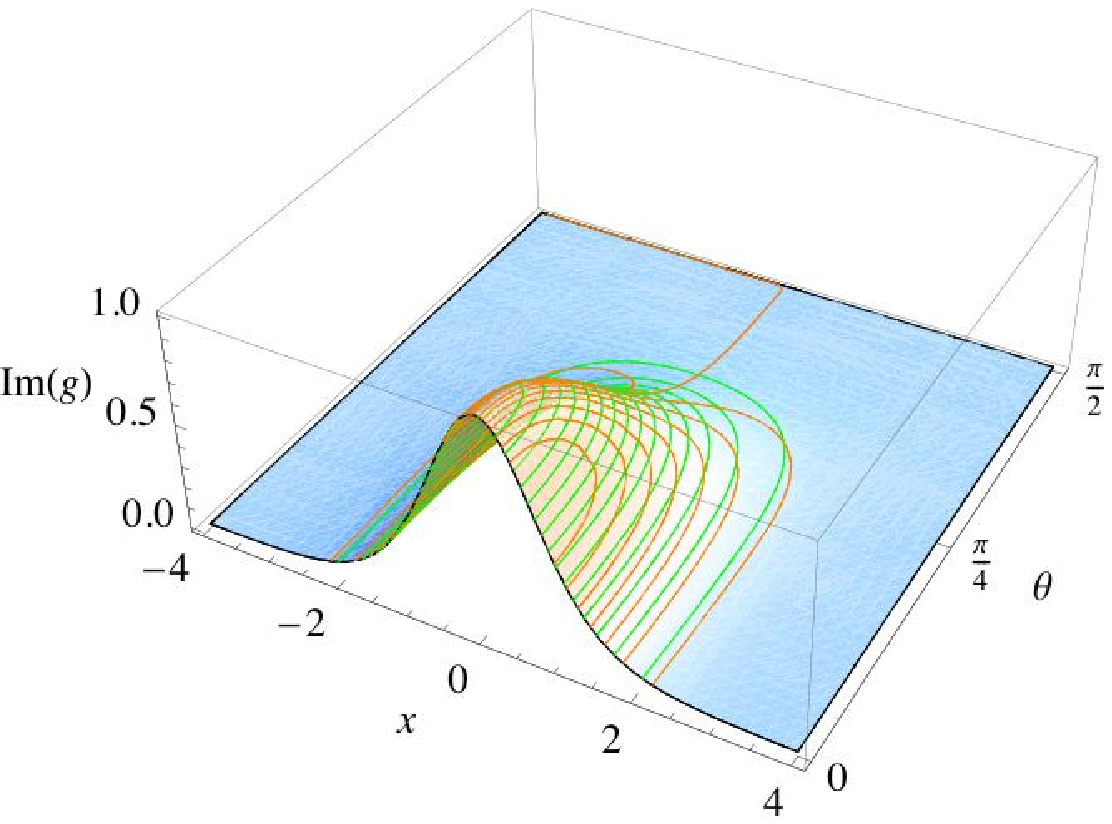}}
\caption{(a) shows the $\re\left[g(x)\right]$ and (b) shows the $\im\left[g(x)\right]$ 
as defined by equation \ref{eq:gx1} with $A=B=1$. The orange and green lines show the contours of 
the real and imaginary parts of $g$ respectively.}
\label{fig:gx1}
\end{figure}

Thermal and quantum fluctuations in a Bose-Einstein condensate cause a small-but-finite population of 
non-condensed particles. When a soliton is present in the 
system these non-condensed particles bunch up in the low-density region around the 
soliton~\cite{ds_law2003,damski2006}. 
Without paying close attention to the 
specific details of this non-condensed density, we assign $g(x)$ [of equation \ref{eq:nonhomogeneous_equation}] 
the following fairly generic form;
\begin{equation}
 g(x)=\cos^4(\theta)\left[A\tanh\left(x_c\right)\sech^2\left(x_c\right)+iB\sech^2\left(x_c\right)\right]
\label{eq:gx1}
\end{equation}
where $A$ and $B$ are real constants [$g(x)$ is shown in Fig. \ref{fig:gx1} with $A=B=1$]. 
Note that we have chosen $g(x)$ to have the same symmetry as $\psi_0$ (that 
is the real part is odd, while the imaginary part is even) and that $g(x)$ decays at the same rate as $1-|\psi_0|^2$. 
As boundary conditions on $\psi_1$ we simply say that $\psi_1(x)\rightarrow$constant and 
$D_x\psi_1(x)\rightarrow0$ as $x\rightarrow\infty$, 
aswell as basic symmetry arguments; $\re\left[\psi_1(x)\right]=-\re\left[\psi_1(-x)\right]$ and 
$\im\left[\psi_1(x)\right]=\im\left[\psi_1(-x)\right]$. 
 
\begin{figure}
\centering
\subfiguretopcaptrue
\subfigure[]{\includegraphics[width=6cm]{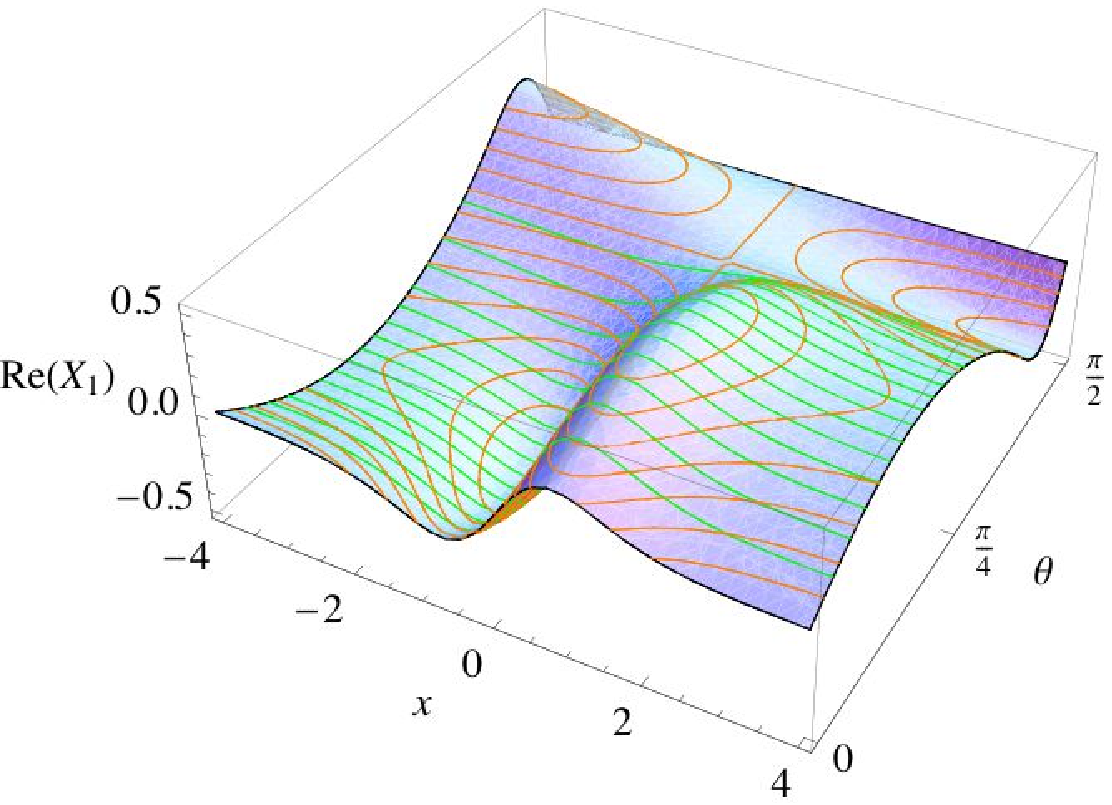}}
\qquad
\subfigure[]{\includegraphics[width=6cm]{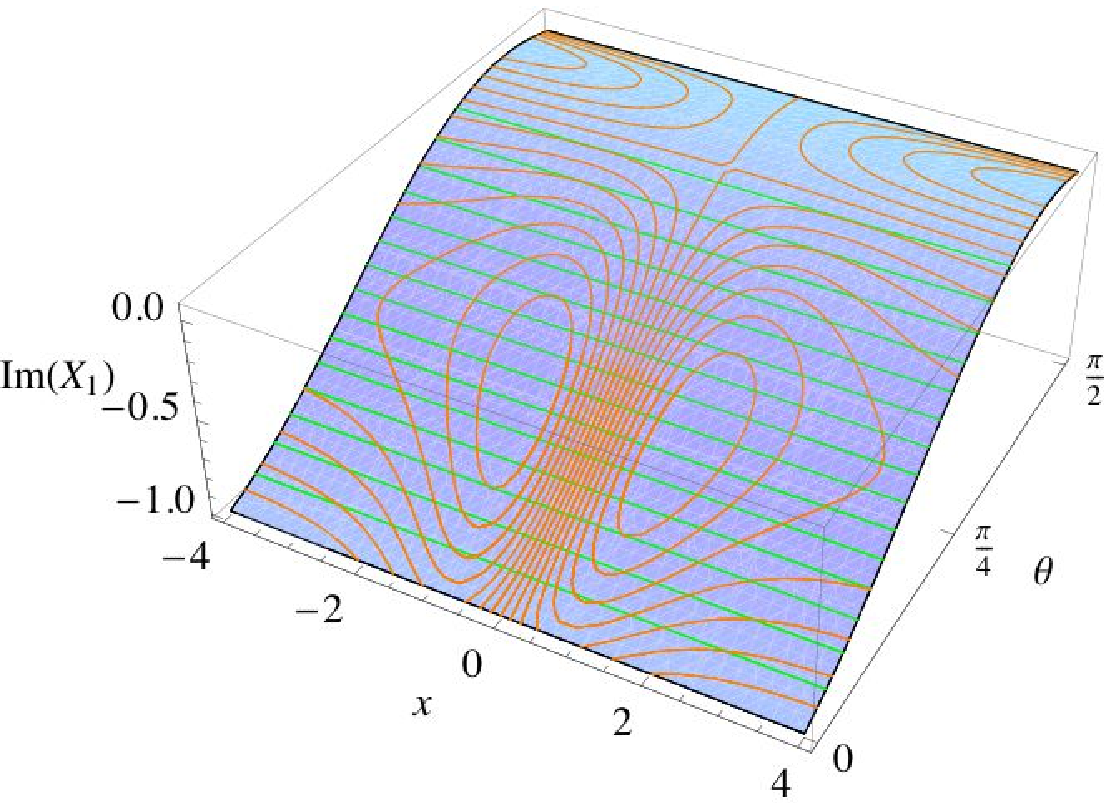}}
\caption{(a) shows the $\re\left[\psi_1(x)\right]$ and (b) shows the $\im\left[\psi_1(x)\right]$ 
as defined by equation \ref{eq:X1bec} with $A=B=1$. The orange and green lines show the contours of 
the real and imaginary parts of $\psi_1$ respectively.}
\label{fig:X1}
\end{figure}

Divergences in $\psi_1$ as $x\rightarrow\infty$ can be avoided by the conditions;
\begin{equation}
 \lambda_1^4=-\kappa_1^4,\quad\lambda_2^4=-\kappa_2^4,\quad\lambda_3^4=-\kappa_3^4,
\end{equation}
and the symmetry is ensured by the conditions;
\begin{equation}
 \lambda_1^2=0,\quad\lambda_1^3=-\frac{1}{4}\sec^2(\theta),\quad\lambda_2^3=-\frac{1}{8}\sec(\theta)\tan(\theta).
\end{equation}
These six additional conditions give us the Greens matrix,
\begin{eqnarray*}\fl
 \tilde{G}_{11}(x>s)=
 \frac{\sec^2(\theta)}{4}\omega_1(x)\bar{\omega}_3(s)+
\frac{\sec(\theta)\tan(\theta)}{8}\omega_2(x)\bar{\omega}_3(s)+
\frac{\sec^3(\theta)}{32}\omega_2(x)\bar{\omega}_4(s)-\nonumber\\
\frac{\sec^2(\theta)}{4}\omega_3(x)\bar{\omega}_1(s)-
\frac{\sec(\theta)\tan(\theta)}{8}\omega_3(x)\bar{\omega}_2(s)-
\frac{\sec^3(\theta)}{32}\omega_4(x)\bar{\omega}_2(s),\\
\fl
 \tilde{G}_{11}(x<s)=
 -\frac{\sec^2(\theta)}{4}\omega_1(x)\bar{\omega_3}(s)-
\frac{\sec(\theta)\tan(\theta)}{8}\omega_2(x)\bar{\omega}_3(s)-
\frac{\sec^3(\theta)}{32}\omega_2(x)\bar{\omega}_4(s)+\nonumber\\
\frac{\sec^2(\theta)}{4}\omega_3(x)\bar{\omega}_1(s)+
\frac{\sec(\theta)\tan(\theta)}{8}\omega_3(x)\bar{\omega}_2(s)+
\frac{\sec^3(\theta)}{32}\omega_4(x)\bar{\omega}_2(s),\\
\fl
 \tilde{G}_{12}(x>s)=
 \frac{\sec^2(\theta)}{4}\omega_1(x){\omega}_3(s)+
\frac{\sec(\theta)\tan(\theta)}{8}\omega_2(x){\omega}_3(s)+
\frac{\sec^3(\theta)}{32}\omega_2(x){\omega}_4(s)-\nonumber\\
\frac{\sec^2(\theta)}{4}\omega_3(x){\omega}_1(s)-
\frac{\sec(\theta)\tan(\theta)}{8}\omega_3(x){\omega}_2(s)-
\frac{\sec^3(\theta)}{32}\omega_4(x){\omega}_2(s),\\
\fl
 \tilde{G}_{12}(x<s)=
 -\frac{\sec^2(\theta)}{4}\omega_1(x){\omega}_3(s)-
\frac{\sec(\theta)\tan(\theta)}{8}\omega_2(x){\omega}_3(s)-
\frac{\sec^3(\theta)}{32}\omega_2(x){\omega}_4(s)+\nonumber\\
\frac{\sec^2(\theta)}{4}\omega_3(x){\omega}_1(s)+
\frac{\sec(\theta)\tan(\theta)}{8}\omega_3(x){\omega}_2(s)+
\frac{\sec^3(\theta)}{32}\omega_4(x){\omega}_2(s),
\end{eqnarray*}
\begin{figure}
\centering
\subfiguretopcaptrue
\subfigure[]{\includegraphics[width=6cm]{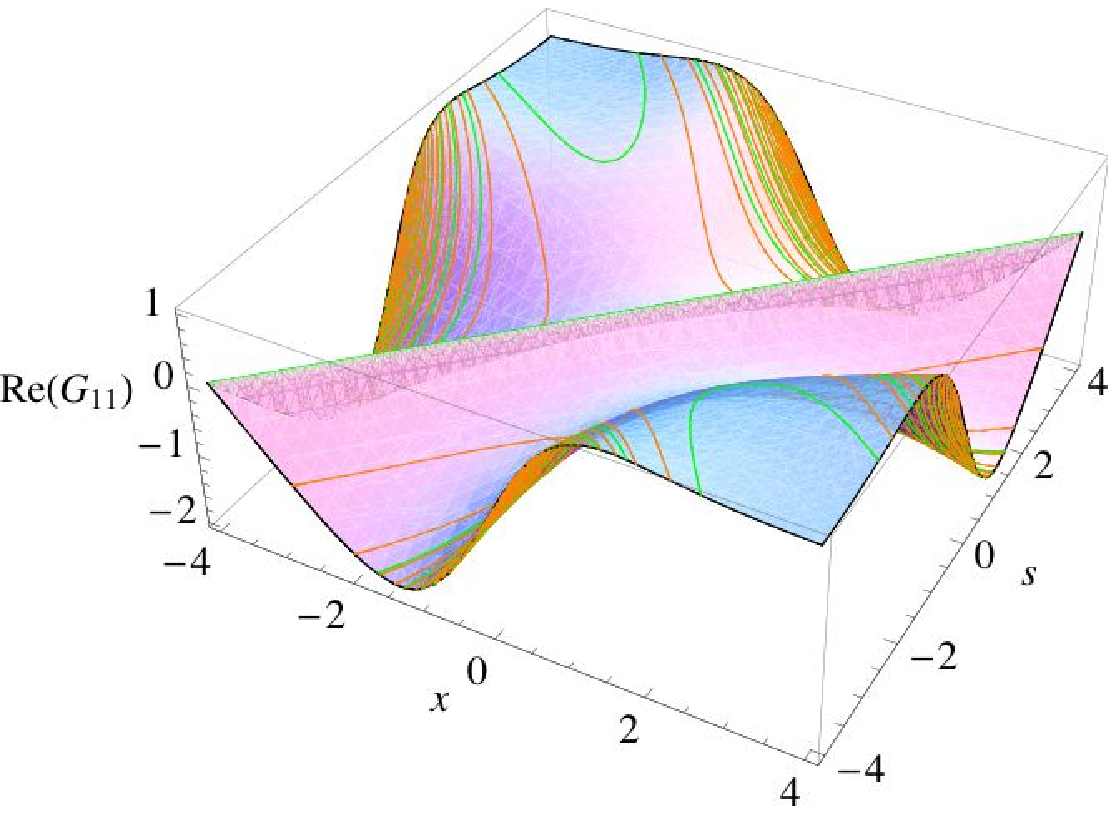}}
\qquad
\subfigure[]{\includegraphics[width=6cm]{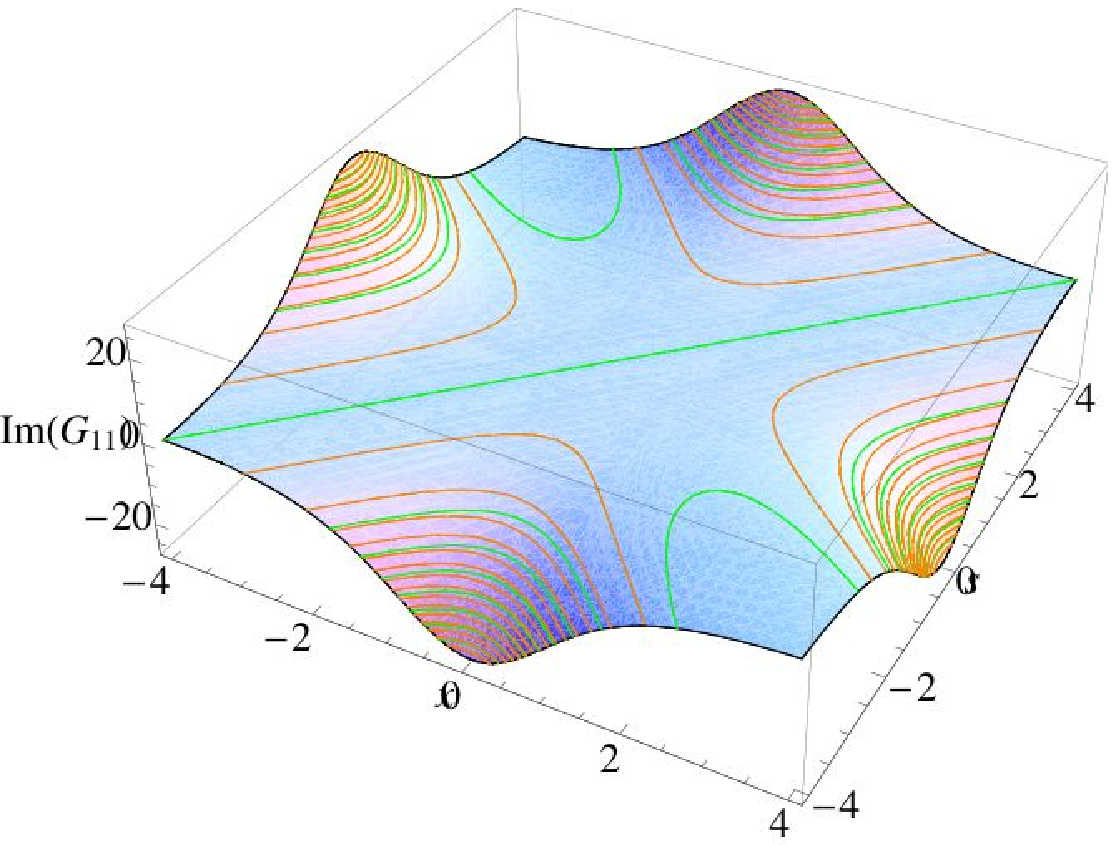}}
\\
\subfigure[]{\includegraphics[width=6cm]{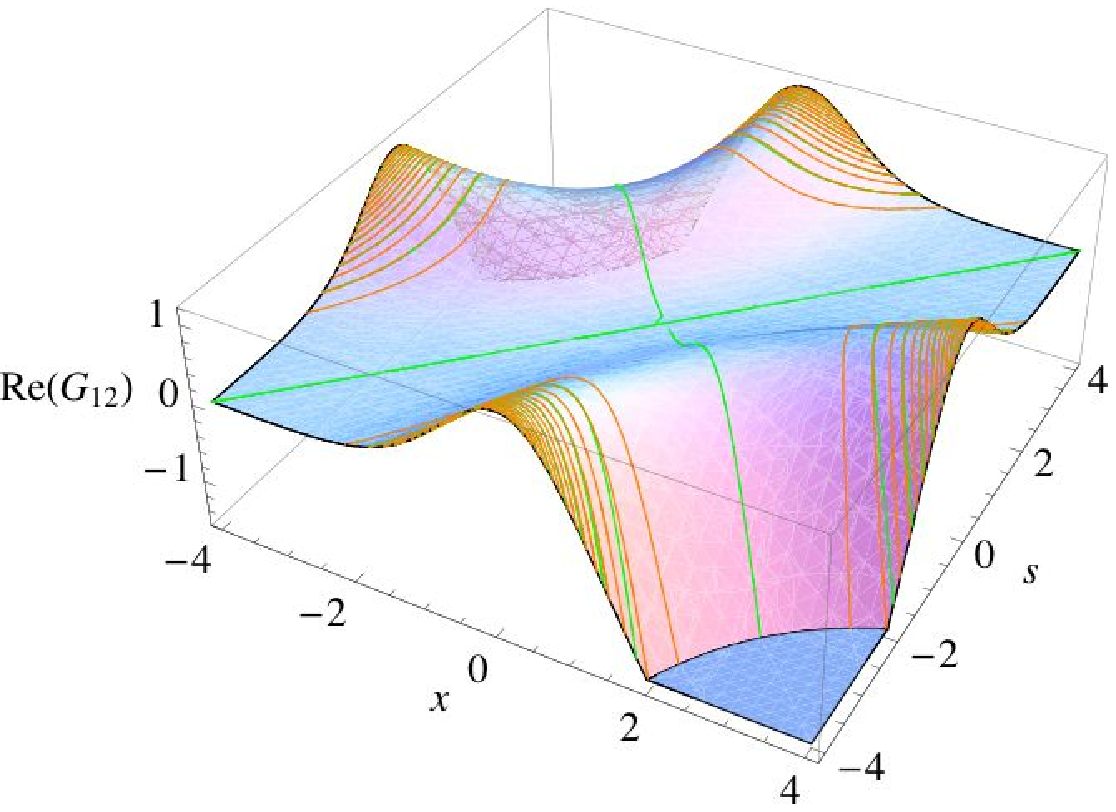}}
\qquad
\subfigure[]{\includegraphics[width=6cm]{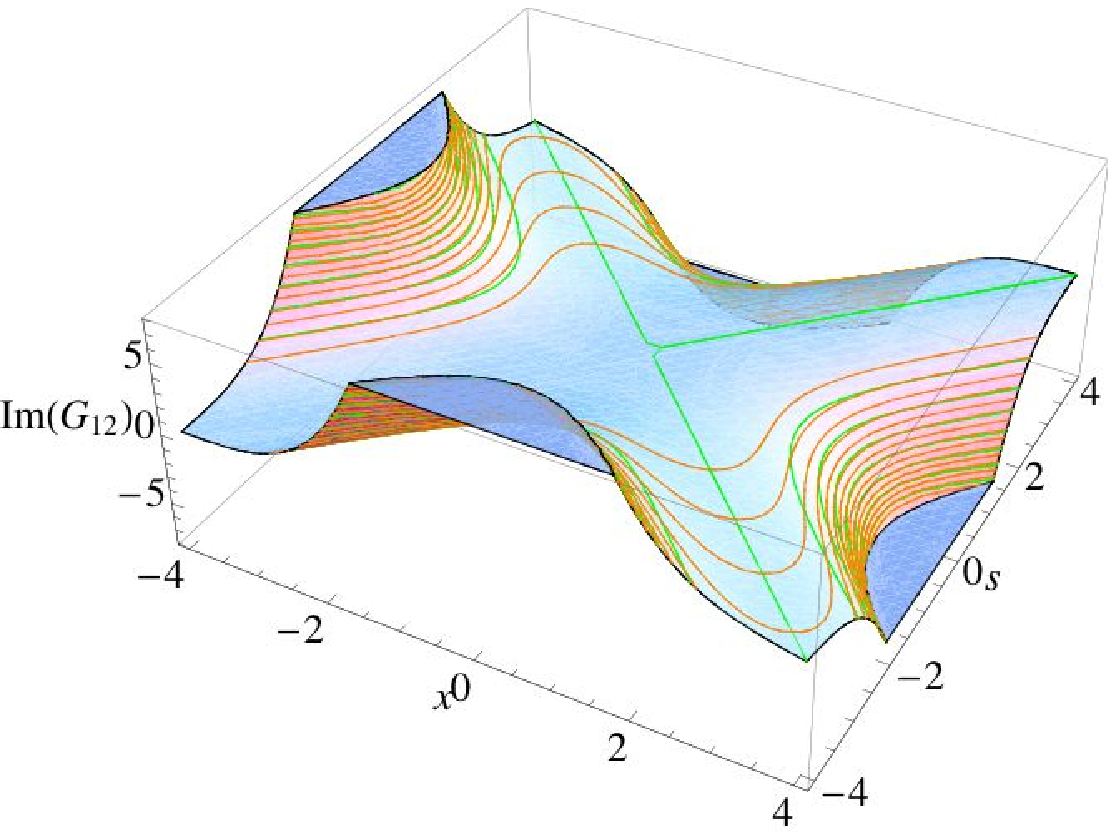}}
\caption{(a) and (b) show the real and imaginary parts of $\tilde{G}_{11}$ 
and (c) and (d) show the real and imaginary parts of $\tilde{G}_{12}$ (we have set $\theta=\pi/4$). 
The orange and green lines show the contours of 
the real and imaginary parts of the function respectively.}
\label{fig:mean_field_solution}
\end{figure}
$\tilde{G}_{21}$ and $\tilde{G}_{22}$ are easily deduced from the symmetry of $\tilde{G}$. 
The expression for $\psi_1$ then follows,
\begin{eqnarray}\fl
\psi_1(x)=\frac{1}{4}\sech^2(x_c)\Big[2x_c\big(A\cos(2\theta)+B\sin(2\theta)\big)+
\sin(\theta)\big(2B\cos(\theta)-\big.\Big.\nonumber\\
\Big.\big. A\sin(\theta)\big)\sinh(2x_c)\Big]+
\frac{i}{2}\cos(\theta)\big[A\sin(\theta)-2B\cos(\theta)\big]\label{eq:X1bec}
\end{eqnarray}
and $\psi_1$ is plotted in Fig. \ref{fig:X1}. One can easily check that equation~\ref{eq:X1bec} is indeed 
a solution to equation~\ref{eq:nonhomogeneous_equation} with $g(x)$ defined by equation~\ref{eq:gx1}.

\section{Conclusion and discussion}

In this article we have introduced four exact analytic solutions 
to the NLS equation linearised around a dark soliton [equation~\ref{eq:zero}]. 
These solutions are given in equations~\ref{eq:om1}--\ref{eq:om4}. 
These four solutions provide a possible means of bypassing the need to solve the spatial perturbative correction (denoted $\psi_1(x)$ 
in this paper) using the complete 
set of finite $E$ eigenfunctions [given in equations~\ref{eq:bdg_solns1a}--\ref{eq:bdg_solns1b}] supplemented with 
generalised eigenfunctions for the nullspace of $\mathcal{H}_x$, (a procedure which appears 
to be common-place in the previous literature in-spite of it's apparent 
difficulty~\cite{bilaspavloff_darksoliton,chinese_bdg_solutions}). 
To illustrate this point, we 
constructed a Green's matrix which can be used to find a solution to 
equation~\ref{eq:nonhomogeneous_equation} once boundary conditions have been defined. 
We applied the technique to the problem of thermal and/or 
quantum fluctuations within a Bose-Einstein condensate. 

It is interesting to note that, of the four solutions presented in equations~\ref{eq:om1}--\ref{eq:om4} 
only two of them [$\omega_1(x)$ and $\omega_2(x)$] remain bounded in the limit as $x\rightarrow\infty$. 
The other two, $\omega_3(x)$ and $\omega_4(x)$, are linearly diverging and exponentially diverging 
respectively. This then begs the question as to which set of perturbing functions [$g(x)$ 
in equation~\ref{eq:nonhomogeneous_equation}] are amenable to the use of the Greens matrix defined by 
equation~\ref{eq:green}, particulary when the boundary conditions require $\psi_1$ to be bounded. 
Certainly in the example problem of Section~\ref{example1} where the 
perturbing function itself is strongly localised around the soliton, satisfying boundary 
conditions does not seem to be an issue, since the integral in equation~\ref{general_solution} 
is able to contain the divergences associated with $\omega_3$ and $\omega_4$. 
It is also possible to contain divergences by exploiting even or odd symmetries of $g(x)$, 
since $\omega_3$ and $\omega_4$ have even and odd symmetries in the real and imaginary parts, 
the integration in equation~\ref{general_solution} can once again, avoid undesired divergences. 
Intuitively one might expect (due to the fact that the only interesting parts of 
equations~\ref{eq:om1}--\ref{eq:om4}  are in the region close to the soliton) 
that any perturbing function which has a considerable nonzero component far away from the 
soliton would require the use of the radiative solutions given in equations~\ref{eq:bdg_solns1a}--\ref{eq:bdg_solns1b},  
and one would follow the procedure of Ref.~\cite{chinese_bdg_solutions}. However, a general 
theory on this issue is currently lacking.

\ack

AGS wishes to thank Alan Bishop, Avadh Saxena, and David Roberts for useful 
discussions. 

\section*{References}

\bibliographystyle{unsrt}
\bibliography{bib_soliton}

\end{document}